\documentclass[a4paper,cameraready]{Interspeech}
\usepackage{kotex}
\usepackage{times}
\usepackage{latexsym}
\usepackage{amsmath}
\usepackage{booktabs}
\usepackage{url}
\usepackage{hyperref}
\usepackage{graphicx}
\usepackage{subcaption}  
\usepackage{float}
\usepackage{multirow}
\usepackage{xcolor}

\title{VorTEX: Various overlap ratio for Target speech EXtraction}

\author[affiliation={1},equalcontribution]{Ro-hoon}{Oh}
\author[affiliation={1},equalcontribution]{Jihwan}{Seol}
\author[affiliation={1},correspondingauthor]{Bugeun}{Kim}

\address{
    $^1$ Department of Artificial Intelligence, Chung-Ang University, Republic of Korea
}

\email{\{heiscold, seoljh0722, bgnkim\}@cau.ac.kr}


\usepackage{comment}

\begin{document}
\maketitle

\begin{abstract}
    
    
    


   Target speech extraction (TSE) aims to recover a target speaker’s voice from a mixture. While recent text-prompted approaches have shown promise, most approaches assume fully overlapped mixtures, limiting insight into behavior across realistic overlap ratios. We introduce VorTEX (Various overlap ratio for Target speech EXtraction), a text-prompted TSE architecture with a Decoupled Adaptive Multi-branch (DAM) Fusion block that separates primary extraction from auxiliary regularization pathways. To enable controlled analysis, we construct PORTE, a two-speaker dataset spanning overlap ratios from 0\% to 100\%. We further propose Suppression Ratio on Energy (SuRE), a diagnostic metric that detects suppression behavior not captured by conventional measures. Experiments show that existing models exhibit suppression or residual interference under overlap, whereas VorTEX achieves the highest separation fidelity across 20–100\% overlap (e.g., 5.50 dB at 20\% and 2.04 dB at 100\%) while maintaining zero SuRE, indicating robust extraction without suppression-driven artifacts.
\end{abstract}

\section{Introduction}

Target Speech Extraction (TSE), which aims to recover a clean voice from mixture speech, has recently drawn significant attention. Specifically, with the advance of large language models, researchers have begun to adopt \textit{text prompts} instead of reference audio \cite{llmtse, audiosep, dgmo, textrolmix, interspeaker2025}. However, existing text-prompted TSE models still mainly focus on fully overlapped conditions, limiting their applicability in real-world scenarios \cite{textrolmix}. 
Meanwhile, audio-prompted TSE models have addressed multiple overlap conditions, providing more realistic robustness across diverse conversational conditions \cite{Zhang24p_OR_TSE, Zhao24_C_TSE}.
Thus, in this paper, we suggest a text-prompted TSE model, VorTEX, which is the first attempt to address wide-range of overlapping conditions. 

Prior TSE models widely assumed two implicit assumptions on their performance on different overlapped scenarios.
First, lower-overlap mixtures are easier to achieve higher performance than fully-overlapped ones, thus a model trained on full overlap should improve as overlap decreases.
Second, a TSE model primarily performs extraction of the target rather than suppression of the mixture.
Though these two assumptions seem obvious, surprisingly, we demonstrate counter-examples with a new synthesized \textit{dataset}, PORTE (Prompts for Overlap-Robust Target speech Extraction), where both assumptions are fail\footnote{Dataset: [Blinded for review]}.
We carefully designed PORTE, following audio-prompted TSE datasets, which already cover various overlap degrees \cite{Zhang24p_OR_TSE, Zhao24_C_TSE}. As current TSE methods were trained and tested on fully-overlapped datasets only, we re-train those TSE models on PORTE and verify whether they satisfy those implicit assumptions. Then, we observe that the performance of prior models are not dramatically better as overlap degree becomes lower on PORTE. Also, some prior TSE models exhibit suppression-like behavior instead of true extraction of target signal from the mixed one, even though PORTE adopts text prompts mirroring simple but mostly plausible usage of text-prompted TSE.

After such an unexpected observation, we aim to search a possible solution that are aligned with the real world.
In the real world, the overlap degree might differ across mixed audio files. This variance makes it unrealistic to switch between different models or manually segment inputs depending on the overlap condition. We suspect a single procedure in computation mechanism as the cause of such unexpected behavior. Existing studies often place most of the burden on a single fusion mechanism, even though the acoustic characteristics of mixtures change substantially with overlap. This seems leading to overlap-specific failure modes, including unstable trade-offs between separation and perceptual quality; to achieve higher quality, they decide to suppress highly overlapped audio. Hence, we propose a design that separates the process of true extraction from the process of regulating general extraction quality. Specifically, in our model, VorTEX (Various overlap ratio for Target speech EXtraction), we introduce a multi-branch structure called Decoupled Adaptive Multi-branch (DAM)\footnote{Code: [Blinded for review]}. DAM combines three mutual-complementary fusion blocks: a main extractor and two regularizer modules. This design enables VorTEX to achieve its best performance regardless overlap degrees and demonstrate the possibility of satisfying two aforementioned assumptions in a single text-prompted TSE model.

Concretely, our main claim is overlap-robustness. And, such robustness cannot be evaluated thoroughly using a single metric. As mentioned above, a good TSE model should first truly separate target signals from the mixture. The quality of produced audio becomes the second important measure. Though prior work usually adopted both signal similarity metrics (e.g., SISDR) and perceptual quality metrics (e.g., PESQ or WER), they failed to satisfy the first condition due to suppression; a mixture-suppressed audio could achieve high signal similarity and high perceptual quality, indicating improvement from the mixed audio. Thus, we propose another kind of measurement that can detect such suppression: Suppression Ratio on Energy (SuRE). Using SuRE and previously used other metrics, we compare VorTEX with other previous benchmark models, showing VorTEX's true extraction behavior. 

In summary, we propose a new VorTEX architecture along with the PORTE dataset to tackle the current mismatch between text-prompted TSE and the real-world requirements. Conducting a comparison experiment and an ablation experiment, we compare VorTEX with previous architecture and demonstrate that VorTEX successfully separates the target signal from the mixture, and others are not. As a result, this paper has the following contributions to the community.
\begin{itemize}
    \item \textbf{PORTE dataset:} Our new dataset reveals previous extractors often suppress highly-overlapped mixtures, thereby failing to truly extract the target signal.
    \item \textbf{VorTEX model:} Our model shows that multi-branch mechanism might help to avoid suppression behavior when extracting signals from the mixture.
    
    \item \textbf{SuRE metric:} We suggest a new metric that discriminates suppression behavior from true ability of extracting signals. 
    
    \item \textbf{Consistently balanced performance:} VorTEX achieves consistently strong SISDR across diverse overlap ratios while avoiding extreme degradation and exhibiting zero SuRE, indicating robust extraction behavior rather than suppression-driven artifacts.
    
\end{itemize}

\section{Related Works}
Our aim is to build a text-prompted TSE model that is robust on various overlap degrees. Thus, this section reviews previous work regarding three aspects: dataset, architecture, and metrics.

Regarding the datasets, text-prompted TSE dataset have not considered various overlap situations. Current text-prompted TSE datasets have mainly focused on fully-overlap scenarios \cite{textrolmix, seki2025languageQueriedTSE}. For example, TextrolMix dataset \cite{textrolmix} synthesized mixed speech by fully overlapping multiple speech in TextrolSpeech \cite{textrolspeech}. Similarly, PromptTSE \cite{seki2025languageQueriedTSE} also generated fully-overlapped situations, though they suggested different kind of prompt descriptions for identifying target speech. However, the need of various overlap scenarios has been emphasized in audio-prompted TSE studies \cite{Chen2020LibriCSS, Hershey2016, cosentino2020librimix}. Researchers proposed various datasets with different overlap degrees since audio-prompted TSE models which are trained with fully-overlap scenarios did not successfully handle unseen partial overlap scenarios \cite{cosentino2020librimix}.
For example, LibriCSS \cite{Chen2020LibriCSS} was designed for partial overlap scenarios (from 0\% to 40\%), inspired by usual overlap ratio observed in conference or sequential dialogue. 
LibriMix and SparseLibriMix \cite{cosentino2020librimix} proposed sampled multiple overlap bins (e.g., 0, 20, 40, 60, 80, 100\%) and evaluated performance per overlap ratio. Thus, we need a new text-prompted TSE dataset which covers different overlap degrees.


Regarding the model architecture, current text-prompted TSE studies have limited discussion on how to explicitly handle different overlap situations. Instead, many systems employ a single computation path \cite{Luo2019ConvTasNet, Sato2024SpeakerBeam, Zeng2025USEF, Ao2024USED}, expecting the same mechanism to generalize across overlap regimes. For example, Conv-TasNet \cite{Luo2019ConvTasNet} performs time-domain separation, and SpeakerBeam \cite{speakerbeam}
integrates speaker conditioning with efficient sequence modeling. Moving toward text-based prompts, StyleTSE \cite{textrolmix} introduces a style-aware encoder to extract prosodic features—such as pitch and emotion—from text, which are then used as a style-based embedding to guide the extraction process. More recently, LLM-TSE \cite{llmtse} has extended this by utilizing the semantic reasoning capabilities of large language models to process complex natural-language instructions for identifying the target speaker.
However, despite these advancements in conditioning methods, a single computation path cannot handle different overlap scenarios properly. Low- and mid-overlap mixtures often require faithful reconstruction of the dominant target speech while avoiding unnecessary separation artifacts, whereas high-overlap mixtures demand strong disentanglement of simultaneous signals. This suggests that text-prompted TSE can benefit from a multi-path fusion design with asymmetric roles: a primary extractor that performs the core separation and auxiliary paths that regularize optimization. 
Beyond fusion design, prior approaches may implicitly work as a role as suppression of non-target speeches, which can improve separation metrics while harming perceptual quality. 
This mismatch motivates evaluating overlap-robust TSE not only by SISDR but also by perceptual and suppression-related metrics (e.g., PESQ and STOI) and analyzing whether the model behaves as an extractor rather than a suppressor.

Regarding the metrics, existing TSE studies typically evaluate performance using commonly adopted separation and perceptual metrics such as SDR \cite{scheibler2022sdr}, SISDR \cite{sisdrloss}, and PESQ \cite{pesq}. However, these metrics have a limitation in that they do not explicitly whether the model performs faithful extraction and suppression of the mixture speech. 
For instance, SISDR can yield high scores even when parts of the signal are aggressively attenuated, as long as the remaining parts align well with the reference in the projection sense. Likewise, PESQ provides only a single scalar score and cannot analytically attribute quality changes to either faithful target reconstruction or suppression-driven attenuation of mixed signals.
This ambiguity becomes problematic under varying overlap ratios. At low overlap intervals, suppressors that simply reduce uncertain sections gain competitiveness, but at high overlap ratios, they may fail to maintain the actual timbre and continuity of the target. 
Recent analysis further suggests that SISDR can be misaligned with perceptual attributes \cite{Jepsen2025ASO}: when references contain noise, SISDR is theoretically bounded by the reference SNR and may incentivize reproducing reference noise in the estimate. The analysis also empirically reported a negative correlation between SISDR and perceived noisiness via non-intrusive quality assessment.
Therefore, when evaluating overlap-robust text-prompted TSE, we need a behavior-centric analysis that clearly distinguishes whether the target is actually extracted or suppressed, alongside existing metrics.

\section{PORTE Dataset}


We construct PORTE (Prompts for Overlap-Robust Target Speech Extraction), a synthetic two-speaker dataset designed to evaluate text-prompted TSE under diverse overlap scenarios. PORTE is generated from LibriTTS \cite{Zen2019_LibriTTS} with four steps.

\textbf{Step 1: Source Sampling and Level Control.}
Two utterances from different speakers are randomly sampled. We removed leading silence and discarded utterances shorter than 5 seconds. Also, to adjust the length of each source speech within 5--10 seconds, we truncated longer ones. Then, we resampled all signals to 16 kHz. Following the construction procedure of TextrolMix dataset \cite{textrolmix}, we normalize each source to a target loudness uniformly sampled from $[-33,-25]$ LUFS. Also, the mixture SNR is drawn from a zero-mean Gaussian distribution with 4 dB standard deviation. 

\textbf{Step 2: Overlap-Controlled Mixture Generation.}
We generated mixtures with different overlap degrees by delaying one source relative to the other to match a predefined overlap ratio. Then we randomly assigned the target speaker and sampled the order of speeches in the speech from "first" or "later." After deciding the order of speeches, we sample one of six overlap ratios borrowed from  \cite{cosentino2020librimix}: 0, 20, 40, 60, 80, and 100\%. Based on the overlap ratios and speech order, we added the signal to produce mixture audio. Especially for the 0\% overlap, we inserted a random pause of 0.5--1.2 seconds between speech signals to avoid back-to-back concatenation.

\textbf{Step 3: Text Prompt Annotation.}
For each mixture, we automatically generate a simple natural-language prompt specifying the target speech. We specified target speech with intuitive attributes observable in the mixture: \emph{speaker's gender} when two speech is a mixture of different genders, \emph{the order of speeches} (first or later), and \emph{relative speaking duration} of target speech compared to the other speech (shorter or longer). We used simple but intuitive prompted templates which might be used by the end users, as shown in Table~\ref{tab:porte_prompt_taxonomy}.

Note that our prompts are simple and constructed exclusively from those three attributes. To make the experiment simple and easy to interpret, we do not use lexical content, speaker identities, or oracle time boundaries when generating prompts. Since gender contrast, speaking order, and relative duration are perceptible from the mixture signal itself, the failure of a model could be attributed to the model itself, instead of the difficulty of the dataset. Also, the characteristics makes us to evaluate the performance of extraction without considering explicit ground-truth alignment or waveform information.

\textbf{Step 4: Data Packaging.}
For each mixture, we store (i) the mixture waveform, (ii) time-aligned zero-padded target and interfering signals, and (iii) metadata including overlap ratio, SNR, LUFS values, start/end timestamps, and prompt type. This structure enables easier supervised training and controlled evaluation across overlap conditions.

As a result, PORTE contains 39K two-speaker mixtures. The length of input mixture audio ranges from 5 to 21.13 seconds. We further split the dataset into 36K training and 3K test samples, considering diverse overlap scenarios.

\begin{table}[t]
  \centering
  \footnotesize
  \setlength{\tabcolsep}{3pt}
  \begin{tabular}{@{}l@{ - }p{0.7\columnwidth}@{}}
    \toprule
    \textbf{Type} & \textbf{Prompt template} \\
    \midrule
    Speaker's Gender & Extract only the \{male/female\} voice from this audio. \\
           & Please remove the \{male/female\} voice from this audio. \\
    Speech Order  & Extract the voice of the speaker who spoke \{first/later\}. \\
    Relative Length & Extract the speech that contains a \{shorter/longer\} duration of speech. \\
    \bottomrule
  \end{tabular}
  \caption{Prompt templates used in PORTE datasets.}
  \label{tab:porte_prompt_taxonomy}
\end{table}

\section{VorTEX Model}
\begin{figure}[t]
    \centering
    \includegraphics[width=\linewidth]{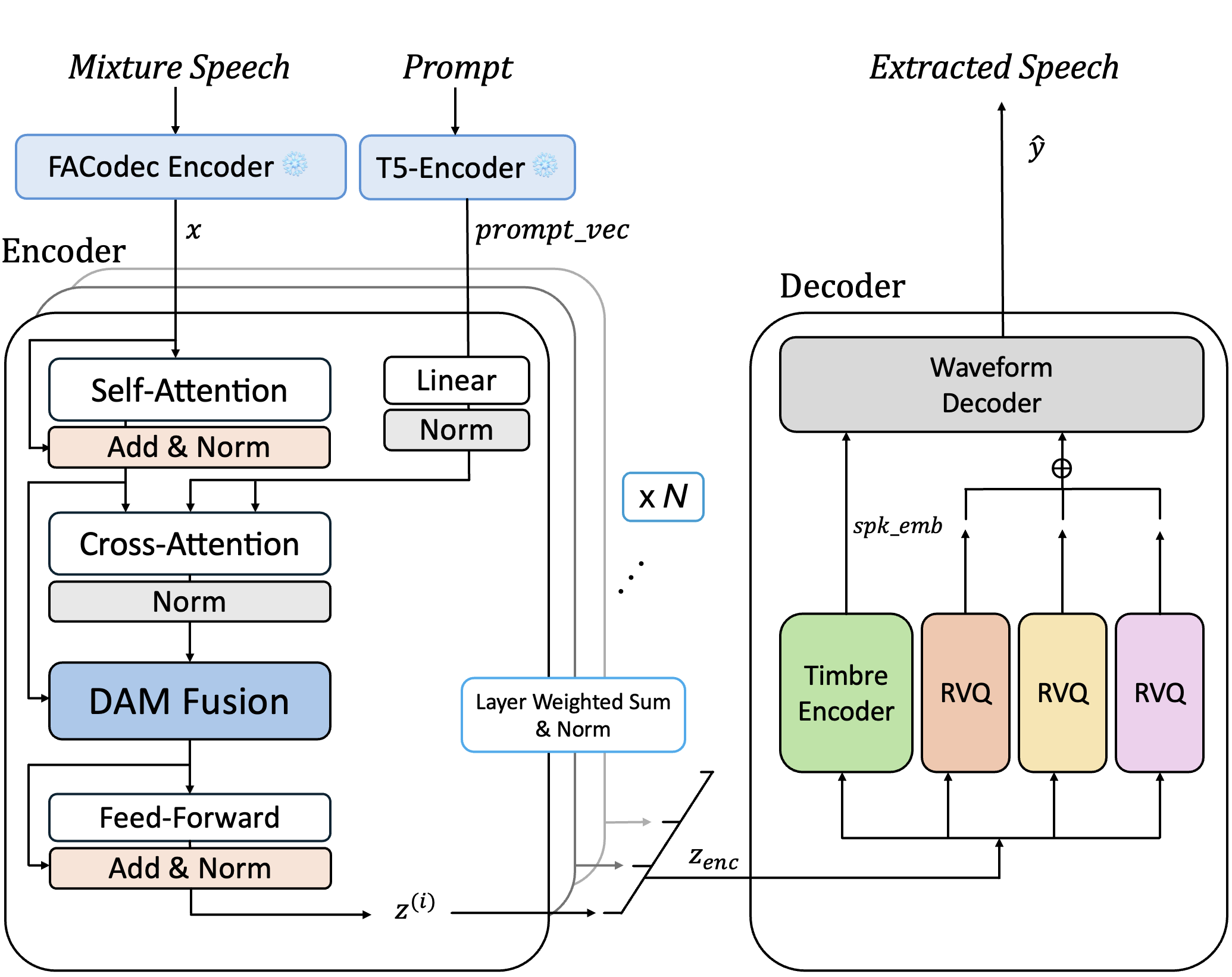}
    \caption{Architecture of VorTEX}
    \label{fig:model}
\end{figure}

\begin{figure*}[t]
    \centering
    \includegraphics[height=5.6cm, keepaspectratio]{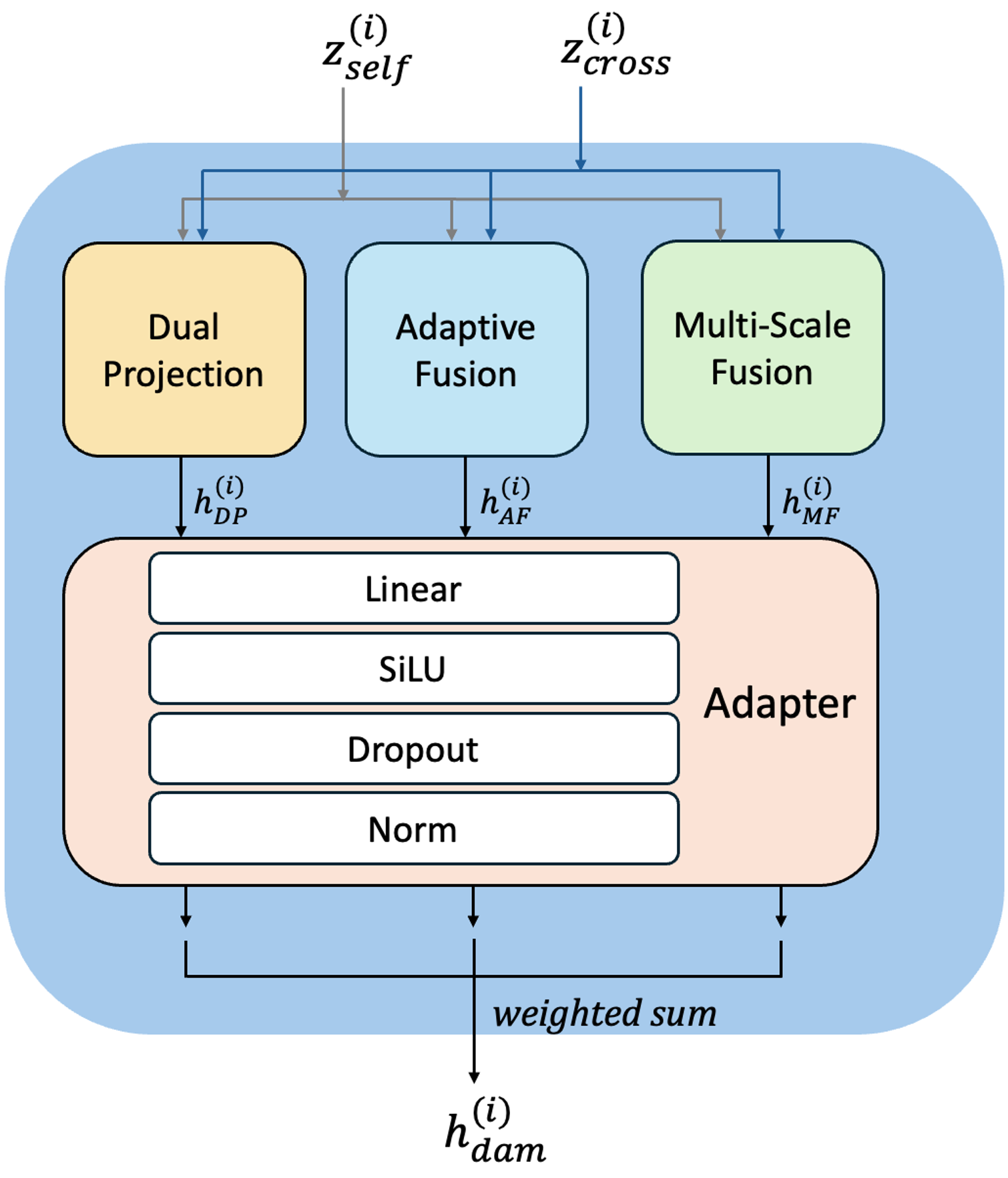}
    \includegraphics[height=5.6cm, keepaspectratio]{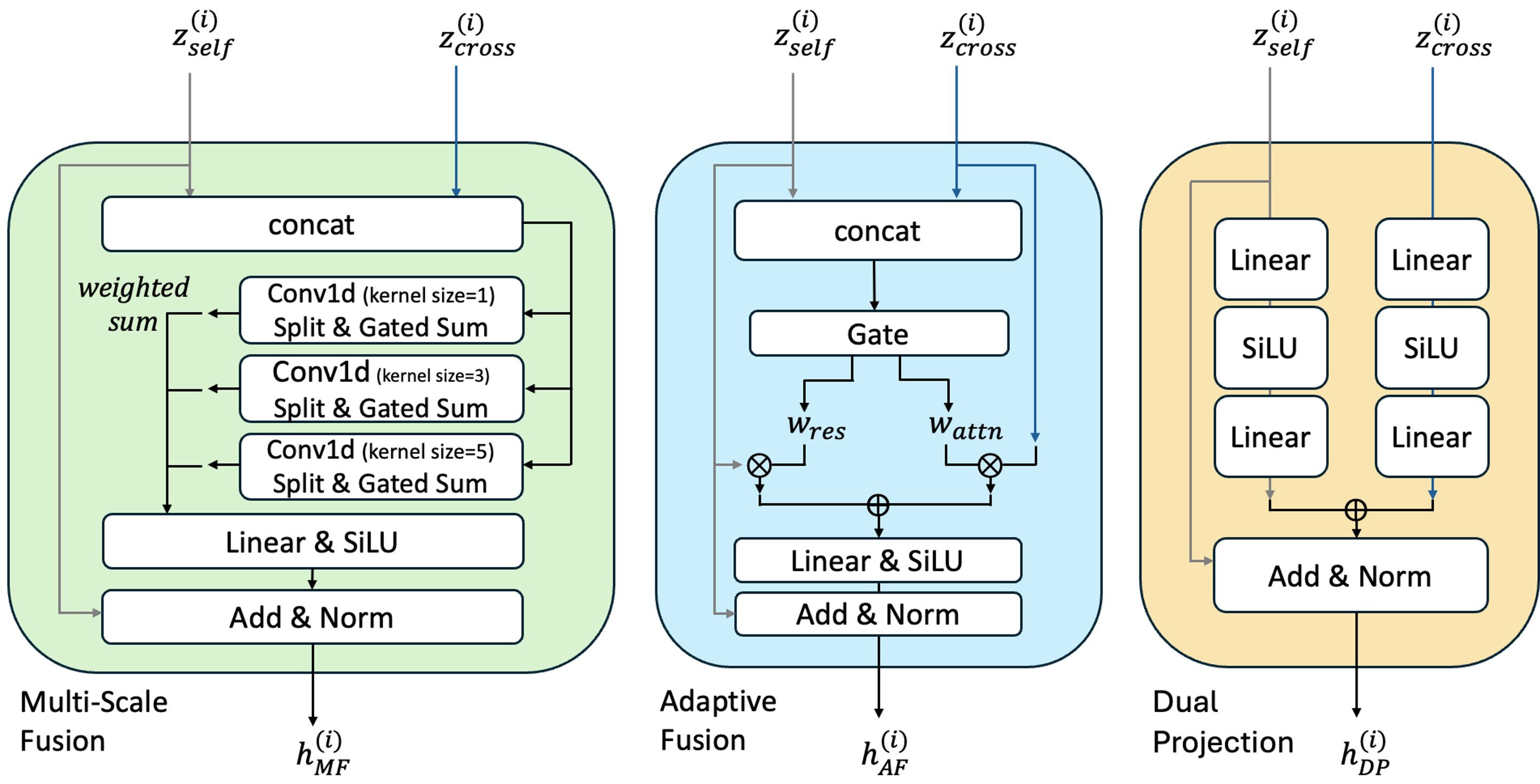}
    \caption{DAM Fusion and its fusion blocks}
    \label{fig:fb}
\end{figure*}

\vspace{0.5em} 





The VorTEX model adopts an Encoder-Decoder architecture, as illustrated in Figure~\ref{fig:model}. First, the input mixture speech is encoded into a speech feature vector $x$ using a frozen speech Encoder. Simultaneously, the natural language prompt is converted into a fixed embedding vector $prompt\_vec$ via a frozen text encoder. Two encoded signals are then fed to the Encoder, which is a variant of Transformer model. The Encoder takes both the speech feature vector $x$ and the prompt vector $prompt\_vec$ as inputs to extract the hidden representation $z_{enc}$ of the target speech. Finally, the Decoder converts $z_{enc}$ into a waveform.

Specifically, we adopted the following pre-trained models to enhance performance. For the speech encoder, we adopted FACodec Encoder \cite{Ju2024NaturalSpeech3}, because it can leverage generalized acoustic representations trained on a large-scale TTS dataset, thereby enhancing both the training efficiency and performance of the TSE model. For the text encoder, we adopted T5-Encoder, specifically \texttt{flan-t5-base}~\cite{chung2022scaling}, since it is widely adopted for effectively vectorizing diverse natural language expressions, ensuring that text-based conditions (prompts) are directly integrated into the speech extraction process. For the Decoder, we initialized the architecture with pretrained FACodec Decoder, with the same reason that we adopted FACodec Encoder. With these pre-trained models, VorTEX can extract speech signal from the mixture. The detailed architecture is described below.




\subsection{Encoder}



The Encoder is based on a Transformer architecture~\cite{vaswani2017attention} which can integrate the conditional relationship between the prompt embedding ($prompt\_vec$) and the speech feature vector ($x$). Basically, computational procedure follows the same procedure of Transformer encoder blocks. 
On $i$-th block, two embedding signals, $z^{(i-1)}$ and $prompt\_vec$, are mixed through a self-attention and a cross-attention layer, and transformed to contextual embedding $z^{(i)}$ using a feed-forward network. 
Here, we define $z^{(0)} := x$ for simplicity in the equation.
And, we should note that we used Rotary Positional Embedding (RoPE) \cite{su2021roformer} instead of standard positional embedding in self-attention to ensure the model can process long input without loss of information.

To ensure the model recognize and handle different overlap degrees, we inserted DAM in each block between cross-attention and feed-forward network. 
The DAM block transforms the cross-attention output $z^{(i)}_{cross}$ using three different pathways. 
Though cross-attention block attempts to identify the target signal from the mixed input using similarity, it may not fully consider the difference between mixed signal and target signal. 
Thus, by juxtaposing identified signal $z^{(i)}_{cross}$ and its original self-attentive signal $z^{(i)}_{self}$, we want to make DAM recognize difference between extracted and original signals in each transformer block.
Mathematically, DAM computes its output $h^{(i)}_{dam}$ with the following formula:
\begin{equation}
    h^{(i)}_{dam} = \sum_{b: \textrm{ block}} w_b\cdot f_b(z^{(i)}_{cross}, z^{(i)}_{self}),
\end{equation}

where $f_b$ and $w_b$ indicates internal pathways for comparing two signals and its corresponding learnable weight. In this paper, we used three pathways, as shown in Figure \ref{fig:fb}: Multi-scale, Adaptive, and Dual Projection.

The first pathway, Multi-scale Fusion, is the main extraction pathway for comparing two signals. To capture multiple periodic scales, we used three 1D convolution layers with different kernel sizes and computed weighted sum of them. Here, we used adaptive weights to make the model utilize difference between two input signals. So, Multi-scale Fusion can automatically weigh significance of multiple periodic differences and predict appropriate contextual representation for the target.


The second pathway, Adaptive Fusion, is an auxiliary fallback pathway for Multi-Scale Fusion. This pathway computes weighted sum of two input signals, for each timestep. Similar to Multi-scale fusion, this utilizes difference between two input signals and compute adaptive weights. So, Adaptive Fusion can provide frame-level comparison, which is simpler and thus less fluctuating than Multi-scale Fusion during the training.


The last pathway, Dual Projection Fusion, is the simplest way for comparing two signals. 
Specifically, $z^{(i)}_{self}$ 
and $z^{(i)}_{cross}$ pass through different MLP layers independently and are combined by a simple summation. 
While this block alone is not sufficient for comprehensive comparison, it could provide a baseline for the comparison. So, using with other pathways, Dual Projection Fusion can improve stability of DAM across different overlap degrees.

Plugging in these three blocks, VorTEX can recognize difference between input and the processed signals. Also, instead of discarding intermediate output $z^{(i)}$, we take a weighted sum across them to preserve the caught information. As a result, we obtain the final encoder representation $z_{enc}$.





\subsection{Decoder}
The Decoder generally follows the FACodec Decoder \cite{Ju2024NaturalSpeech3}. 
The Decoder receives predicted target signal $z_{enc}$ and produces RVQ codes and timbre embedding. Using those codes and timbre embedding, a waveform decoder reconstructs speech $\hat{y}$.

The primary motivation of using FACodec Decoder is representational compatibility with our encoder, which is trained to extract disentangled, target-conditioned representations under varying overlap conditions. 
Specifically, FACodec decodes factorized attribute streams (e.g., content, prosody, and acoustic details) into waveform space, making it suitable for reconstructing structured latent representations. 
However, our Decoder is not intended to directly decode such attributes separately; rather, we wanted to the model consider such aspects during the generation procedure. 
Thus, we decided to borrow the architecture and initial pre-trained parameters of FACodec Decoder, instead of adopting its training pipeline\footnote{When we followed loss functions used in \cite{Ju2024NaturalSpeech3}, including gradient reversal layers and prediction heads, the training of VorTEX becomes highly instable. This is another practical reason why we excluded loss dedicated to each attributes.}.



Moreover, we also adopted the rotation trick \cite{Fifty2024Restructuring} to the residual vector quantization components\footnote{\url{github.com/lucidrains/vector-quantize-pytorch}} to stabilize the training procedure. 
The rotation trick reshapes the gradient passed through the quantization operation and can improve training stability of RVQ, enhancing codebook utilization in practice and reduce collapse risk.

\subsection{Loss functions}
Our training objective $\mathcal{L}$ combines three loss functions: $\mathcal{L}_{\textrm{SISDR}}$, $\mathcal{L}_{\textrm{spk}}$, and $\mathcal{L}_{\textrm{commit}}$. 
Here, $\mathbf{y}$ denotes the ground-truth target speech.
\begin{eqnarray*}
    \mathcal{L}
    &=& \mathcal{L}_{\textrm{SISDR}}
    + \lambda_{s }\mathcal{L}_{\textrm{spk}} + \lambda_{c}\mathcal{L}_{\textrm{commit}}, \\
    \mathcal{L}_{\textrm{SISDR}} &=& -10\log_{10}\left(\frac{\left\lVert \frac{\hat{\mathbf{y}}^{\top}\mathbf{y}}{\left\lVert \mathbf{y}\right\rVert^{2}}\mathbf{y}\right\rVert^{2}}{\left\lVert \frac{\hat{\mathbf{y}}^{\top}\mathbf{y}}{\left\lVert \mathbf{y}\right\rVert^{2}}\mathbf{y}-\hat{\mathbf{y}}\right\rVert^{2}}\right), \\
    \mathcal{L}_{\textrm{spk}} &=& \mathcal{L}_{\textrm{Huber}} + 0.5 \mathcal{L}_{\textrm{cosine}}.
\end{eqnarray*}

First, the SISDR loss $\mathcal{L}_{\textrm{SISDR}}$ computes similarity between ground truth $\mathbf{y}$ and the estimated one $\hat{\mathbf{y}}$. Using this loss function, we wanted to optimize scale-invariant waveform reconstruction under varying overlap conditions.

Second, the speaker embedding loss $\mathcal{L}_{\textrm{spk}}$ enforces speaker consistency between the extracted signal and the reference target speech. 
Specifically, the reference speaker embedding is extracted from the ground-truth target speech $\mathbf{y}$ using the FACodec encoder and the speaker/timbre embedding network provided in the FACodec decoder. 
In parallel, we estimate a speaker embedding from the estimated output waveform $\hat{\mathbf{y}}$. 
Then, we compute the discrepancy between the reference and predicted embeddings using a combination of Huber loss and cosine similarity loss, which stabilizes regression while encouraging directional alignment in the embedding space.

Third, the commitment loss $\mathcal{L}_{\textrm{commit}}$ is implemented using an RVQ module and corresponds to the commitment term. 
Concretely, it is defined as an MSE loss between the pre-quantization latents (inputs to RVQ) and the corresponding quantized latents, where the quantized path is detached. 
This encourages continuous latents to commit to their selected quantized representations, stabilizing quantization dynamics and reducing representation drift during discrete latent learning.

For the weighting coefficients, we set them as $\lambda_s = 5$ and $\lambda_c = 0.05$. Those values are set heuristically to balance reconstruction fidelity, speaker alignment, and quantization regularization during joint optimization.

\section{Experimental Setup}

\subsection{Evaluation Metric}
To compare the performance of TSE approaches, we used five evaluation metrics: SISDR, SISDRi, PESQ, WER, and SuRE. The first two metrics evaluate separation fidelity, or how well the model separated target signal from the mixed input. Specifically, SISDR (Scale-Invariant Signal-to-Distortion Ratio) \cite{sisdrloss} measures the absolute separation quality of the extracted speech. This quantifies residual distortion between the reference target signal and the model output under scale-invariant conditions. And, SISDRi (SISDR improvement) 
measures the relative gain over the input mixture, indicating how much a model improves 
extraction compared to the input. 


The next two metrics measure perceptual quality, or whether the extracted signal is intelligible. Specifically, PESQ (Perceptual Evaluation of Speech Quality) \cite{pesq} evaluates perceptual speech quality by estimating how natural the extracted speech sounds. And, WER (Word Error Rate) measures transcription accuracy and serves as an ASR-based proxy for linguistic correctness. In the experiment, we obtained transcripts using Whisper-v3 \cite{whisper} and computed WER accordingly. 

However, these two metrics cannot identify whether a TSE model extracts a signal or falsely manipulates input mixture. Thus, we suggest a new metric, Suppression Ratio on Energe (SuRE). 
SuRE is a sort of false negative rate of energy activation; it measures the proportion of target-active frames in which the estimated signal energy is substantially suppressed relative to the ground-truth energy, thereby providing an indicator for suppression.
Given the ground-truth target speech $y$ and the estimated speech $\hat{y}$, both signals are first cropped according to the boundaries where speaker in $y$ starts and ends to speak. Then, we compute RMS energy per frame; namely, $g_i$ for $y$ and $\hat{g}_i$ for $\hat{y}$ for $i$-th frame. Then, we compute SuRE as follows:
\begin{equation}
    \text{SuRE} = \frac{\sum_i \mathbb{I}[(g_i > \tau) \land (\hat{g}_i <\beta g_i)]}{\sum_i\mathbb{I}[g_i > \tau]},
\end{equation}
where $\mathbb{I}(\cdot)$ is an indicator function whose value is one if and only if the given condition satisfied. We used two parameters $\tau$ and $\beta$ for deciding whether the frame has speech signal or suppressed. First, $\tau$ is an peak-relative 
threshold for detecting silence in the ground truth speech. We set $\tau=0.01 \cdot \max \{g_i\}$. Second, $\beta$ is a relative threshold for detecting suppression in the estimated speech. We set $\beta = 0.1$, corresponding to substantial energy suppression in signals by 20 dB.




\subsection{Comparison Experiment}

To test the performance of VorTEX, we conduct a comparative experiment using testset of our PORTE dataset. 
Specifically, we evaluate VorTEX against four models. Two are prominent models in the broader audio source separation (ASS) task, which are often used as baselines in TSE studies: AudioSep \cite{audiosep} and DGMO \cite{dgmo}. And, other two are text-prompted TSE models: StyleTSE \cite{textrolmix} and LLM-TSE \cite{llmtse}. These models were selected not only for their public availability but also for their foundational roles and distinct architectural approaches that serve as rigorous benchmarks for text-guided extraction. The following paragraphs elaborates each model.

AudioSep is a foundation model for universal source separation, designed to separate any sound based on natural language queries. It serves as a critical baseline because it establishes a standardized framework for text-to-audio alignment, which is directly analogous to the core challenge of text-prompted TSE. 

DGMO, conversely, represents the state-of-the-art in high-fidelity source separation. It employs a discriminative learning objective that highlights target while suppressing interference through an optimized time-frequency masking mechanism. 

StyleTSE is a representative model for attribute-based target speech extraction, which introduces a style-aware encoder that maps text cues into a style-based embedding. This embeddings includes detailed prosodic features such as pitch and emotion, which acts as a conditioning signal to guide TSE model. 

LLM-TSE, on the other hand, represents the state-of-the-art in instruction-based TSE through the integration of Large Language Models (LLMs). It leverages the advanced semantic reasoning ability of LLMs to interpret complex, natural-language queries that may describe target speakers in various conversational contexts.

\subsection{Ablation Experiment}
Additionally, to assess the contribution of each fusion-related component to the overall performance of VorTEX, we conduct a controlled ablation study. 
We compare the full model with several variants in which one fusion component is removed at a time, including Multi-Scale Fusion (MF), Dual Projection (DP), Adaptive Fusion (AF), and the DAM Fusion block.
Each ablation variant differs from the full model only by removing the corresponding component, while all other architectural and training configurations remain unchanged. 
In particular, the encoder–decoder backbone, the text encoder and prompt-conditioned pathway, the training dataset, the optimization schedule, and the evaluation protocol are kept identical. 

Through this controlled ablation setting, we aim to quantify the contribution of each fusion-related component and to clarify how these components affect feature aggregation and integration under different overlap conditions.
Specifically, we test the following hypotheses. First, by comparing VorTEX with its variant without DAM, we investigate whether the decoupled multi-branch routing and dynamic aggregation of heterogeneous fusion pathways are essential for overlap-robust feature integration. Without DAM’s weighted combination of MF, AF, and DP, we expect more pronounced degradation in extraction fidelity under severe overlap, indicating that structured multi-branch integration is critical for maintaining robustness across diverse mixture conditions.

Second, by comparing VorTEX with its variant without MF, we examine whether the multi-scale gated comparison mechanism serves as the primary extraction pathway that captures periodic and local differences between $z^{(i)}_{self}$ and $z^{(i)}_{cross}$, across overlap ratios, especially under high-overlap conditions where fine-grained disentanglement is required. Since MF employs multiple 1D convolutions with different receptive fields and adaptive scale weighting, we hypothesize that removing MF will significantly reduce extraction fidelity (SISDR), even if perceptual metrics (PESQ) remain relatively stable.

Third, by comparing VorTEX with its variant without AF, we evaluate whether a lightweight frame-wise adaptive weighting mechanism contributes auxiliary regularization effects by providing a simpler and less fluctuating comparison path between residual and cross-attended features. Since AF derives frame-level adaptive weights from the residual and cross-attended features without multi-scale convolution, we expect its removal to cause moderate degradation under overlap, while leaving perceptual smoothness largely unchanged.
Lastly, by comparing VorTEX with its variant without DP, we assess whether independently projected residual and cross-attended streams provide a stabilizing baseline comparison that supports consistent integration across overlap ratios. Since DP performs separate MLP transformations followed by summation, we hypothesize that its removal will produce only marginal changes in overall extraction performance, reflecting its role as a secondary structural complement rather than a core extractor.
\color{black} 

\subsection{Training Configuration}
All experiments were conducted on a machine equipped with three NVIDIA RTX 4090 GPUs using distributed data parallel training. Mixed precision training with bfloat16 (bf16) was employed to improve memory efficiency and computational throughput.
VorTEX was trained for up to 50 epochs with a per-GPU batch size of 1 and gradient accumulation steps of 32, resulting in an effective batch size of 32 per GPU. 
We used the AdamW optimizer with an initial learning rate of $1.5 \times 10^{-4}$ and weight decay of $5 \times 10^{-5}$. The optimizer hyperparameters were set to $\beta_1 = 0.9$, $\beta_2 = 0.98$, and $\epsilon = 1 \times 10^{-9}$. 
For learning rate scheduling, we adopted a linear warm-up strategy for the first 15\% of total training steps, starting from 1\% of the base learning rate, followed by cosine annealing decay down to 5\% of the initial learning rate ($7.5 \times 10^{-6}$). The total number of training steps was computed considering gradient accumulation.
Training was performed for a maximum of 50 epochs with early stopping based on validation loss. Specifically, training was terminated if the validation loss did not improve for 40 consecutive epochs, and the checkpoint with the best validation performance was selected for evaluation.

\begin{table}[!t]
    \small
    \centering
    \begin{tabular}{l|r|r@{\ }r@{\ }r@{\ }r@{\ }r@{\ }r}
    \toprule
      \textbf{(1) SISDR $\uparrow$} & Avg. & 0\% & 20\% & 40\% & 60\% & 80\% & 100\% \\
    \midrule
    AudioSep & 1.03 & \textbf{13.01} & 2.47 & -0.52 & -1.55 & -2.83 & -4.37 \\
    DGMO & -4.56 & -2.92 & -4.29 & -4.52 & -5.06 & -5.30 & -5.25 \\
    StyleTSE & 0.24 & 2.45 & -0.22 & -0.18 & -0.27 & -0.18 & -0.16 \\
    LLM-TSE & -0.51 & -0.51 & -0.71 & -0.55 & -0.47 & -0.43 & -0.40 \\
    VorTEX & \textbf{4.50} & 7.07 & \textbf{5.50} & \textbf{4.53} & \textbf{4.29} & \textbf{3.56} & \textbf{2.04} \\
    \midrule
       \multicolumn{8}{l}{\textbf{(2) SISDRi $\uparrow$}} \\
    \midrule
    AudioSep & -16.79  & -77.84 & -4.70 & -4.83 & -4.20 & -4.35 & -4.81 \\
    DGMO & -21.41 & -91.92 & -9.39 & -7.71 & -7.06 & -6.59 & -5.82 \\
    StyleTSE & -1.78 & -10.42 & -0.05 & -0.01 & -0.07 & -0.05 & -0.09 \\
    LLM-TSE & -0.45 & -0.45 & -0.56 & -0.43 & -0.41 & -0.43 & -0.43 \\
    VorTEX & \textbf{4.56} & \textbf{7.13} & \textbf{5.65} & \textbf{4.66} & \textbf{4.35} & \textbf{3.57} & \textbf{2.01} \\
    \midrule
       \multicolumn{8}{l}{\textbf{(3) PESQ $\uparrow$}} \\
    \midrule
    AudioSep & 1.71 & 3.02 & 1.80 & 1.52 & 1.42 & 1.29 & 1.21 \\
    DGMO & 1.23 & 1.53 & 1.32 & 1.17 & 1.13 & 1.11 & 1.11 \\
    StyleTSE & \textbf{2.13} & \textbf{3.26} & \textbf{2.41} & \textbf{2.07} & \textbf{1.85} & \textbf{1.67} & \textbf{1.52} \\
    LLM-TSE & 1.47 & 2.23 & 1.64 & 1.38 & 1.25 & 1.17 & 1.12 \\
    VorTEX  & 1.26 & 1.31 & 1.28 & 1.26 & 1.24 & 1.24 & 1.24 \\
    \midrule
       \multicolumn{8}{l}{\textbf{(4) WER $\downarrow$}} \\
    \midrule
    AudioSep  & 53.8 & 31.7 & 48.6 & 57.9 & 57.1 & 60.9 & 66.7 \\
    DGMO & 71.1 & 67.7 & 73.8 & 71.8 & 70.9 & 69.8 & 72.8 \\
    StyleTSE & \textbf{35.1} & \textbf{21.9} & \textbf{29.7} & \textbf{31.8} & \textbf{34.5} & \textbf{37.4} & \textbf{55.3} \\
    LLM-TSE & 91.1 & 109.2 & 107.7 & 91.4 & 83.3 & 76.3 & 80.0 \\
    VorTEX & 45.8 & 28.4 & 37.8 & 42.1 & 48.2 & 52.6 & 65.6 \\
    \midrule
       \multicolumn{8}{l}{\textbf{(5) SuRE $\downarrow$}} \\
    \midrule
    AudioSep & 0.38 & 0.83 & 0.64 & 0.44 & 0.25 & 0.09 & 0.06 \\
    DGMO & 0.36 & 0.59 & 0.50 & 0.42 & 0.31 & 0.22 & 0.15 \\
    StyleTSE & 0.00 & 0.00 & 0.00 & 0.00 & 0.00 & 0.00 & 0.00 \\
    LLM-TSE & 0.00 & 0.00 & 0.00 & 0.00 & 0.00 & 0.00 & 0.00 \\
    VorTEX & 0.00 & 0.00 & 0.00 & 0.00 & 0.00 & 0.00 & 0.00 \\
    \bottomrule
    \end{tabular}
    \caption{Result of model comparison experiment}
    \label{tab:result_baseline}
\end{table}

\section{Result and Discussion}
\label{sec:result}

In this section, we analyze the experimental results from both model comparison (Table \ref{tab:result_baseline}) and ablation study (Table \ref{tab:result_ablation}). Rather than focusing on a single score, we examine how performance varies across overlap ratios and highlight characteristic limitations of different models. Table \ref{tab:result_baseline} presents a comparison between VorTEX and baseline models, while Table 3 \ref{tab:result_ablation} analyzes the contribution of individual architectural components. From these results, three main observations emerge: (i) perceptual quality and extraction fidelity exhibit a measurable trade-off across models; (ii) VorTEX maintains relatively robust extraction performance as overlap increases; and (iii) ablation results clarify the primary extraction mechanism and auxiliary regularization-like components.



\subsection{Perceptual Quality versus Separation Fidelity}

First, we examined how perceptual quality (PESQ, WER) metrics relate to waveform-level separation fidelity (SISDR/SISDRi). Table \ref{tab:result_baseline} summarizes model performance across overlap ratios. 
Across models, PESQ and WER do not consistently track SISDR and SISDRi under overlap, indicating that perceptual quality may be partially decoupled from target separation fidelity in interfered mixtures.

We could observe this decoupling when we applied audio source separation model to text-prompted TSE.
AudioSep achieved high PESQ in the non-overlapping condition (3.02), but under overlapped condition, its SISDR dropped sharply (becomes negative beyond 40\%), whereas PESQ and WER degraded more gradually. 
For example, PESQ decreased to 1.80 at 20\% overlap and further to 1.21 at 100\% overlap, while WER increased from 31.7 at 0\% overlap to 66.7 at 100\% overlap. 
DGMO showed a similar trend with negative SISDR across overlap while PESQ and WER change within a narrower range. 
Specifically, PESQ decreased from 1.53 at 20\% overlap to 1.11 at 100\% overlap, and WER increased from 67.7 to 72.8 over the same range. 
These observations reveal a clear decoupling between PESQ/WER and SISDR under overlap.

Besides, previous text-prompted TSE models, StyleTSE and LLM-TSE, showed a different form of mismatch in which perceptual quality remained competitive despite limited separation performance. 
StyleTSE achieved the highest average PESQ of 2.13 and the lowest average WER of 35.1 among the compared models. 
However, its SISDR remained close to zero for higher overlap ratios, even though PESQ and WER remained competitive. 
At 100\% overlap, StyleTSE recorded PESQ 1.52 and WER 55.3, and LLM-TSE recorded PESQ 1.12 and WER 80.0, while their SISDR remained near zero or negative. 
Though the tendency of mismatch differs from that in audio source separation models, these highlight that text-prompted TSE models also suffer from a performance mismatch.

In contrast, VorTEX exhibited a distinct behavior under interference. 
VorTEX achieved the highest SISDR/SISDRi under overlap, while maintaining comparable perceptual/ASR scores across overlap ratios. 
For instance, VorTEX showed SISDR of 4.50 on average, which was higher than other models by 3.5-9.0. Also, its PESQ was 1.28 at 20\% overlap and 1.24 at 100\% overlap, and its WER increased from 37.8 at 20\% overlap to 65.6 at 100\% overlap. 
This suggests that VorTEX improved separation fidelity under interference more effectively than the baselines, without an extreme sacrifice in overall perceptual quality.

\subsection{Result of Model Comparison}
\label{sec:sisdr_comparison}

\begin{figure}[t]
  \centering    
  \includegraphics[width=0.95\columnwidth]{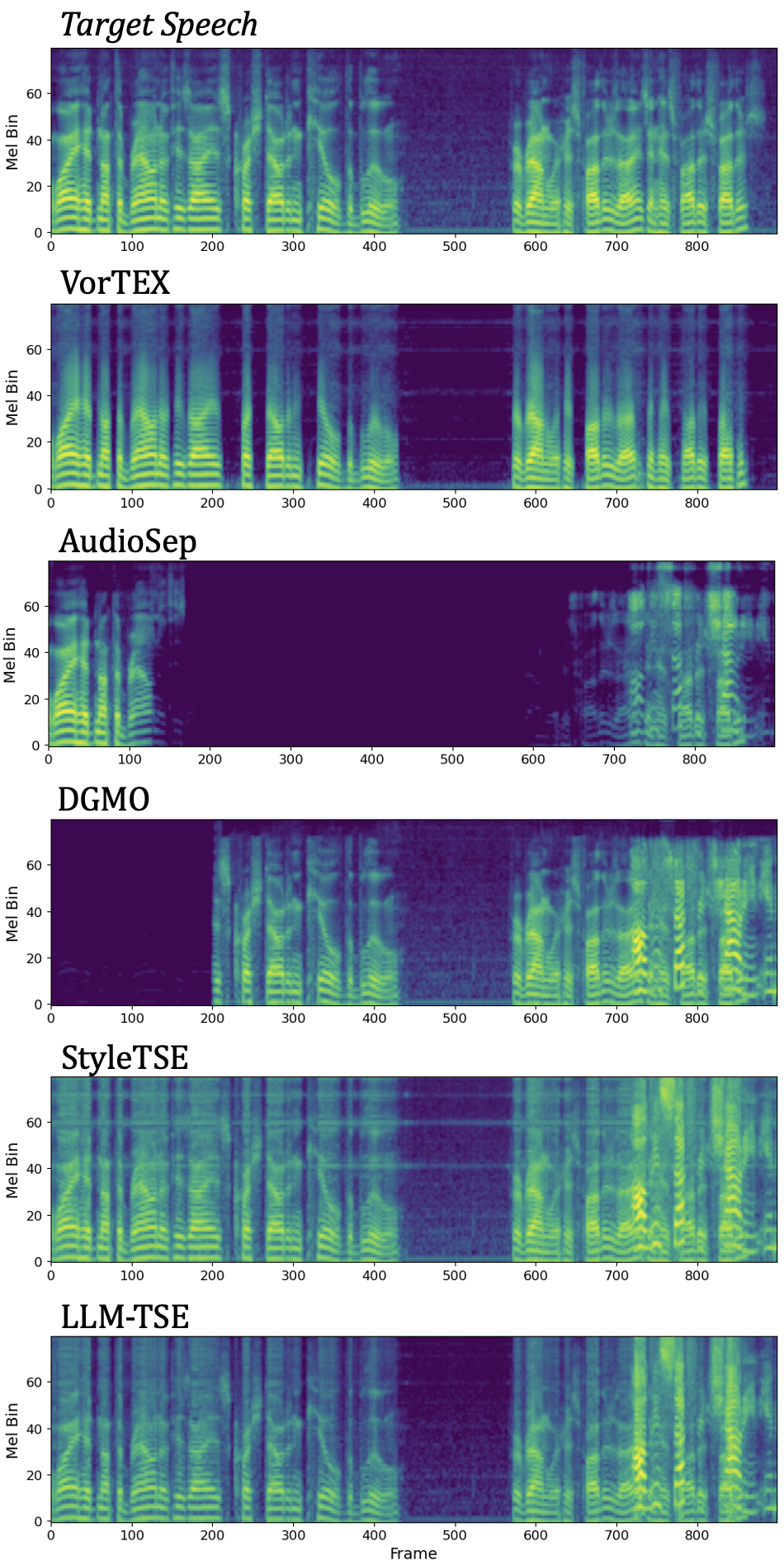}
  \caption{Example mel-spectrograms}
  \label{fig:suppression_mel}
\end{figure}

As mentioned above, existing models suffer performance mismatch. We suspect one possible cause of such mismatch is suppression behavior. So, to highlight such behavioral differences, we compare audio source separation models with text-prompted TSE models regarding SuRE and SISDR. Also, we visualize these suppression patterns with an instance from our PORTE testset; Fig.~\ref{fig:suppression_mel} compares mel-spectrograms of the ground-truth target and extracted outputs.

Audio source separation models, AudioSep and DGMO, showed suppression-associated behavior. AudioSep achieved strong extraction in the clean condition (SISDR 13.01~dB at 0\%), but its SISDR dropped to 2.47~dB at 20\% overlap and becomes negative as overlap increases, reaching to $-1.55$~dB at 60\% overlap. 
Moreover, such degradations co-occurred with non-zero SuRE in target-active regions (0.83 at 0\%, 0.64 at 20\%, and 0.44 at 40\%).
Such suppression behavior could be observed in Figure \ref{fig:suppression_mel}, where AudioSep produced almost empty frames from 150 to 700 in the given sample. 
Similarly, DGMO exhibited consistently negative SISDR under overlap, measuring $-4.29$~dB at 20\% overlap and $-5.06$~dB at 60\% overlap. 
The model also showed non-zero SuRE values (0.59 at 20\% and 0.31 at 60\%).
Figure \ref{fig:suppression_mel} also depicts that DGMO generated empty frames from 0 to 200.
Thus, we can conclude that AudioSep and DGMO's reduced extraction fidelity under overlap is accompanied by suppression effects.
%

In contrast, previous text-prompted TSE models, StyleTSE and LLM-TSE, showed residual interference instead of suppression. Though these models maintained near-zero SuRE across all overlap ratios, their target speech extraction improvements remain limited regarding SISDR. 
StyleTSE achieved a SISDR of 2.45~dB at 0\% overlap, but drops to $-0.22$~dB at 20\% overlap and remained near zero at higher overlap ratios. 
Similarly, LLM-TSE stayed near zero or slightly negative across all conditions (e.g., $-0.51$~dB at 0\% overlap and $-0.40$~dB at 100\% overlap), showing minimal variation as overlap increases. 
One possible cause of lower SISDR score could be found in Figure~\ref{fig:suppression_mel}. While StyleTSE and LLM-TSE avoided suppression-like energy dropouts, residual interference remained in frames from 700 to 1000. 
These observations suggest that these models are limited by insufficient target isolation rather than suppression.

Unlike both groups above, VorTEX showed no suppression and higher separation fidelity. The model maintained SuRE at 0.00 across all conditions while consistently achieving the highest SISDR and SISDRi for overlap ratios of 20\% and above, reaching 5.50~dB at 20\%, 4.29~dB at 60\%, and 2.04~dB at 100\%. 
Accordingly, in Figure \ref{fig:suppression_mel}, harmonic continuity and spectral energy are more consistently preserved across the same mel-frame regions where previous models suffered from suppression or interference. 
Thus, VorTEX achieved stronger extraction performance without introducing suppression-like artifacts nor interference, distinguishing it from previous models.


\begin{table}[!t]
    \small
    \centering
    \begin{tabular}{l|r|r@{\ \ }r@{\ \ }r@{\ \ }r@{\ \ }r@{\ \ }r}
    \toprule
    
       \textbf{(1) SISDR} $\uparrow$ & Avg. & 0\% & 20\% & 40\% & 60\% & 80\% & 100\%\\
    \midrule
    VorTEX & \textbf{4.50} & \textbf{7.07} & 5.50 & 4.53 & \textbf{4.29} & \textbf{3.56} & 2.04 \\
     
    \quad w/o DAM & 3.11 & 6.20 & 4.55 & 2.77 & 2.63 & 1.83 & 0.71 \\
    \quad w/o MF & 3.46 & 5.60 & 4.78 & 2.87 & 3.60 & 2.69 & 1.24 \\
    \quad w/o AF & 4.41 & 7.00 & \textbf{5.55} & 4.52 & 4.01 & 2.96 & \textbf{2.42} \\
    \quad w/o DP & 4.46 & 7.00 & 5.53 & \textbf{4.69} & 4.08 & 3.16 & 2.35 \\
    
    \midrule
    
    
       \multicolumn{8}{l}{\textbf{(2) PESQ} $\uparrow$} \\
    \midrule
    VorTEX  & \textbf{1.26} & 1.31 & 1.28 & 1.26 & 1.24 & \textbf{1.24} & \textbf{1.24} \\

    \quad w/o DAM & 1.26 & 1.31 & 1.28 & 1.26 & \textbf{1.25} & 1.23 & 1.20 \\
    \quad w/o MF & 1.26 & \textbf{1.32} & \textbf{1.29} & \textbf{1.28} & 1.24 & \textbf{1.24} & 1.23 \\
    \quad w/o AF & 1.26 & 1.30 & 1.27 & 1.26 & \textbf{1.25} & \textbf{1.24} & \textbf{1.24} \\
    \quad w/o DP & 1.25 & 1.29 & 1.28 & 1.24 & 1.23 & 1.22 & 1.22 \\

    
    
    \bottomrule
    \end{tabular}
    \caption{The result of ablation study}
    \label{tab:result_ablation}
\end{table}

\subsection{Result of Ablation Experiment}

Table~\ref{tab:result_ablation} analyzes the contribution of individual components of VorTEX to target speech extraction performance under varying overlap conditions. 
Although the architecture follows a parallel fusion design, the ablation results indicate that the modules contributed differently to overall extraction effectiveness, as reflected by SISDR trends across overlap ratios.

First, removing the DAM module lead to notable degradation, particularly under severe overlap. 
The average SISDR decreases from 4.50~dB to 3.11~dB. 
While the reduction at 0\% overlap is moderate (from 7.07~dB to 6.20~dB), the performance gap is widen as overlap increases. 
At 40\% overlap, SISDR decreases from 4.53~dB to 2.77~dB, and at 100\%, from 2.04~dB to 0.71~dB. 
PESQ remained within a similar range (e.g., 1.26 for the full model versus 1.25 without DAM at 80\% overlap), suggesting that the main effect of DAM removal is on separation fidelity rather than perceptual quality. 
These indicate that adopting three pathways might structurally complement to each other to enhance separation performance.

Second, removing the Multi-Scale Fusion (MF) module also resulted in a consistent and substantial reduction in extraction performance. 
The average SISDR decreased from 4.50~dB in the full VorTEX model to 3.46~dB without MF. 
At 0\% overlap, SISDR decreased from 7.07~dB to 5.60~dB. 
At 40\% overlap, the value dropped from 4.53~dB to 2.87~dB, and at 100\%, from 2.04~dB to 1.24~dB. 
Interestingly, PESQ showed slight increases in some low-to-mid overlap conditions (e.g., from 1.28 to 1.32 at 20\% overlap and from 1.26 to 1.28 at 40\%), consistent with the earlier observation that perceptual scores may improve even when extraction fidelity decreases. 
These patterns indicate that MF serves as the primary extraction pathway responsible for isolating target speech under interference.

Third, in contrast, removing Adaptive Fusion (AF) resulted in moderate but limited degradation. 
The average SISDR decreased from 4.50~dB to 4.41~dB, with a more noticeable reduction at higher overlap (e.g., from 3.56~dB to 2.96~dB at 80\%). 
However, PESQ remained nearly unchanged across conditions (e.g., 1.24 for the full model versus 1.24 without AF at 80\%). Thus, we suspect that AF contributes auxiliary regularization-like effects without substantially affecting perceptual metrics.

Lastly, removing Dual Projection (DP) similarly produced only marginal changes in extraction performance. 
The average SISDR decreased slightly from 4.50~dB to 4.46~dB. 
Across overlap ratios, differences remain small (e.g., 5.50~dB versus 5.53~dB at 20\% overlap). 
PESQ differences were similarly minor (1.24 for the full model versus 1.22 without DP at 80\% overlap). Thus, DP could be another type of auxiliary regularizer, maintaining extraction capacity or perceptual quality.

Thus, we conclude that three pathways in DAM module form a complementary relationship among them to fulfill separation behavior.
Multi-Scale Fusion constitutes the core extraction component. And, Dual Projection together with Adaptive Fusion introduce secondary regularization effects that refine extraction behavior without materially altering perceptual scores.

\section{Conclusion}
In this paper, we examined text-prompted TSE under various overlapped conditions, rather than being confined to a fully-overlapped condition. 
Since previous text-prompted TSE models have less considered fully-overlapped condition, models used for text-prompted TSE often suffer from suppression or interference, which is not a solid extraction behavior. Hence, we presented PORTE, a controlled dataset which enables to investigate a model's performance on a wide range of overlap ratios. Also, we proposed VorTEX, an text-prompted TSE architecture centered on Decoupling Adaptive Multi-branch (DAM) Fusion. We trained other models and VorTEX on PORTE dataset and conducted model comparisons and ablation studies using SuRE metric, measuring suppression behavior of a model. Our results indicate that overlap diversity should be treated as an explicit evaluation condition in text-prompted TSE, as model behavior and failure modes could differ.

A central empirical finding is that common intuitions do not always hold in text-prompted extraction. Performance does not necessarily improve as overlap decreases, and some systems can achieve seemingly competitive objective scores by suppressing mixture regions rather than faithfully extracting the target signal. Moreover, as overlap varies, separation fidelity (SISDR/SISDRi) can diverge from perceptual quality metrics, indicating that achieving robustness under overlap involves both strong extraction and careful behavioral diagnostics. In contrast, in our evaluations VorTEX achieves the highest SISDR and SISDRi across 20--100\% overlap conditions while maintaining consistently low SuRE, suggesting improved extraction fidelity without suppression-driven artifacts.

Also, our results suggest that multi-branch fusion mechanism in VorTEX help mitigate these challenges. Ablation results show that removing DAM leads to a clear degradation in extraction performance, particularly in high-overlap conditions, while perceptual metrics remain relatively stable. So, incorporating complementary fusion pathways within a single architecture is associated with improved robustness under the varying overlap conditions evaluated in this study.

Despite these improvements, several limitations remain. First, PORTE is synthetically constructed; therefore, it could be meaningful to conduct a future work validating our findings on real conversational recordings with diverse acoustics, background noise, and multi-party interactions. Second, our prompts rely on a limited set of mixture-observable attributes (e.g., gender, speaking order, and relative duration); an extension study could help the community to widen the prompt space toward richer and more compositional instructions, including speaking style, emotion, and pitch-related cues. Third, although SuRE complements SISDR by diagnosing suppression-like behavior, no single metric fully captures extraction fidelity, perceptual quality, and speaker preservation simultaneously; accordingly, another future work is needed to explore objectives and diagnostics that better preserve target-speaker identity and perceptual naturalness while maintaining high extraction fidelity (SISDR/SISDRi) under various overlaps.

Overall, our findings suggest that overlap variability reveals distinct extraction behaviors that remain less visible under fully-overlapped benchmarks. By explicitly analyzing extraction performance across controlled overlap ratios and incorporating suppression-aware diagnostics, this work highlights the importance of behavior-centric evaluation in text-prompted TSE. 
We hope that our work can help enable a more realistic assessment of text-prompted TSE under overlap variability.

\section{Acknowledgements}
This work was conducted while the author Jihwan Seol was at ELU Lab, Chung-Ang University. The author is currently with NAVER Cloud Corp., Seongnam, Republic of Korea.

\section{Use of Generative AI Tools}

Generative AI tools were used solely for language refinement and grammar correction. 
All technical content, experimental design, and analysis were developed by the authors.

\bibliographystyle{IEEEtran}
\bibliography{mybib}



\end{document}